\documentclass[pre,superscriptaddress,twocolumn,final,showpacs,showkeys,nobibnotes,titlepage,twoside,10pt]{revtex4-1}
\usepackage{amsmath}
\usepackage{graphicx}
\usepackage{amsfonts}
\usepackage{amssymb}
\usepackage{subfigure}
\usepackage{float}
\usepackage{color}
\newcommand{\com}[1]{\textcolor{black}{#1}}

\begin{document}

\author{M. T. Dang}
\affiliation{Van der Waals-Zeeman Institute, University of Amsterdam, The Netherlands}
\author{D. Denisov}
\affiliation{Van der Waals-Zeeman Institute, University of Amsterdam, The Netherlands}
\author{B. Struth}
\affiliation{Deutsches Elektronen-Synchrotron, Hamburg, Germany}
\author{A. Zaccone}
\affiliation{Statistical Physics Group, Department of Chemical Engineering and Biotechnology, University of Cambridge, New Museums Site, Pembroke Street, CB2 3RA Cambridge, UK}
\author{P. Schall}
\affiliation{Van der Waals-Zeeman Institute, University of Amsterdam, The Netherlands}

\begin{abstract}
\com{The mechanical response of glasses remains challenging to understand. Recent results indicate that the oscillatory rheology of soft glasses is accompanied by a sharp non-equilibrium transition in the microscopic dynamics. Here, we use simultaneous x-ray scattering and rheology} to investigate the reversibility and hysteresis of the sharp sharp symmetry change from anisotropic solid to isotropic liquid dynamics observed in the oscillatory shear of colloidal glasses [D. Denisov, M. T. Dang, B. Struth, A. Zaccone, and P. Schall, Sci. Rep. {\textbf{5} 14359 (2015)}]. We use strain sweeps with increasing and decreasing strain amplitude to show that, in analogy to equilibrium transitions, this sharp symmetry change is reversible and exhibits systematic frequency-dependent hysteresis. Using the non-affine response formalism of amorphous solids, we show that these hysteresis effects arise from frequency-dependent non-affine structural cage rearrangements at large strain. These results consolidate the first-order like nature of the oscillatory shear transition and quantify related hysteresis effects both via measurements and theoretical modelling.
\end{abstract}

\title{Reversibility and hysteresis of the sharp yielding transition of a colloidal glass under oscillatory shear}
\maketitle

\section{Introduction}

\com{The flow and relaxation of glasses is important for a wide range of materials including metallic glasses, polymer- and soft glasses, but remains challenging to understand. Glasses are structurally frozen liquids with relaxation times exceeding the experimental time scale by many orders of magnitude and hence exhibiting solid-like properties~\cite{Ediger1996}. Their time-dependent response to  mechanical probing is central to many applications in advanced material science and engineering~\cite{Greer1995,Schroers2009,Greer2010,Argon-book}, but remains a major challenge. Besides applications, insight into this response promises also a deeper understanding of the glassy state as it addresses the important fundamental question of how the arrested glass state respond to an externally applied stress imposing an independent time scale.}

\com{So far the majority of rheological studies has focused on bulk flows, and on stress-strain relationships of glassy materials \cite{Larson-book}. Research on soft glasses such as colloids, emulsions and foams provides growing insight into the microscopic mechanism behind the mechanical response. Recently, these studies reveal interesting signatures of underlying non-equilibrium transitions in the response of glasses to applied shear~\cite{Chikkadi14,Denisov15,Nagamanasa2014,Bartolo2014,Cipelletti2014,Berthier2015}. The idea is that the applied shear provides an external field, to which microscopic rearrangements can couple; coupling also occurs between the rearrangements themselves, mediated by their elastic interactions, leading to long-range correlations in the microscopic deformation of slowly deformed glasses ~\cite{Goyon2008,OlssonTeitel,maloney_robbins09,Lemaitre,Chikkadi11,Chikkadi12,Rahmani2013,Heussinger09,Chikkadi2015}. Colloidal glasses have played an important role in directly visualizing these microscopic correlations. The particles exhibit dynamic arrest due to crowding at particle volume fractions above $\phi_g \sim 0.58$, the colloidal glass transition ~\cite{Pusey1986,vanMegen1998}, and they have been used extensively as models for glasses.}

\com{Under continuous shear, colloidal glasses show long-range correlations in their flow~\cite{Chikkadi11,Chikkadi12}, and have been shown recently to exhibit a first-order transition in the dynamics in response to increasing applied shear rate~\cite{Chikkadi14}.}

\com{A particularly useful technique to investigate the time-dependent response of soft glasses is oscillatory rheology. By measuring the stress response to sinusoidal strain, one determines the linear elastic and viscous moduli $G^\prime$ and $G^{\prime \prime}$ from the harmonic in and out of phase response, respectively, all the way from the linear to the non-linear response regime~\cite{Mason95,Sollich,Hebraud,Fuchs2010}. These measurements provide insight into the yielding ~\cite{Petekidis2002,Petekidis2003,Petekidis2008}, caging and relaxation processes of glasses~\cite{Rogers2011b,Rogers2012, vanderVaart2013} as a function of strain amplitude and frequency. The sinusoidal strain avoids the continuous accumulation of strain and allows frequency-dependent steady states~\cite{Petekidis2013} to be probed. Combined with direct measurement of the particle dynamics, these oscillatory measurements revealed the existence of a critical strain at which particle displacements become irreversible~\cite{Hebraud,Petekidis2002,Bartolo2014,Cipelletti2014,Nagamanasa2014}, as also recently observed in simulations~\cite{Berthier2015}. Independently, our own x-ray scattering measurements during the oscillatory rheology of a colloidal glass showed a sharp symmetry change from anisotropic solid to isotropic liquid-like response~\cite{Denisov15} at the rheological yielding of the glass, suggesting a non-equilibrium first-order transition under the applied oscillatory shear.}

\com{While the relation between these studies needs further investigation, the central question is whether in analogy to first-order equilibrium transitions this sharp transition is reversible, and there are any delay or hysteresis effects.} Following the increasing oscillatory amplitude by a cycle with decreasing amplitude, does the transition, if at all reversible, occur at the same critical strain? Addressing these questions would provide important insight into this non-equilibrium transition. In the oscillatory rheology, hysteresis effects could arise due to structural changes of the glass that alter its rigidity and mechanical properties. For example in the liquid state, restructuring due to "shear thinning" can lower the viscosity, enabling the glass to flow with less dissipation. Moreover, subtle structural rearrangements can also occur in the solid state; the resulting structural changes would lead to pronounced hysteresis effects depending on the frequency of probing.

In this paper, we investigate the reversibility and hysteresis of the sharp non-equilibrium transition using simultaneous x-ray scattering and oscillatory rheology on colloidal glasses. We find that the sharp symmetry change is indeed reversible and exhibits systematic frequency-dependent hysteresis. Using our structural order parameter, we can for the first time identify solid and liquid states of the sheared glass apart, and measure hysteresis in the transition between them during strain amplitude sweeps. Our rheological measurements indicate that this hysteresis results from restructuring due to "shear thinning" in the liquid regime at high strain. The hysteresis almost vanishes for volume fractions around the glass transition, but becomes pronounced for densities above and below. By employing a mean-field model based on the non-affine response formalism for amorphous solids~\cite{Zaccone11, Zaccone11b, Zaccone13, Zaccone13b}, we identify this non-monotonous behavior as due to the competition of dissipation and finite-rate non-affine motion leading to entropic changes of the material. These results provide deeper insight into the nature of this non-equilibrium first-order transition in oscillatory sheared colloidal glasses.

\section{Experiment}
The colloidal glass consists of silica particles with a diameter of $50$ nm and a polydispersity of $10\%$ to prevent crystallization. The particles are suspended in water with a small amount ($1 mM$) of $NaCl$ to screen the particle charges. The Debye screening length is $2.7$ nm, resulting in an effective particle diameter of $2r_0 = 55.4$ nm. Dense samples around the colloidal glass transition were prepared by diluting centrifuged samples. The effective volume fractions estimated assuming a sediment volume fraction of $64\%$ are
$\phi \sim 56\%, 58\%, 59\%$ and $60\%$ $\pm0.5\%$. Investigation of the continuous shear rheology~\cite{Denisov2013,Amann15} showed extended shear-thinning regimes typical for colloidal glasses, and consistent with mode-coupling theory predictions~\cite{Amann15}. For the $\phi=58\%$ sample, these measurements yielded a relaxation time of $\tau \sim 10^6 t_B$~\cite{Denisov2013}, with $t_B$ the relaxation time at infinite dilution, consistent with this estimated volume fraction~\cite{Ackerson1991}. To measure the rheology and structure factor simultaneously, we placed an adapted commercial rheometer (Mars II, Thermo Fisher) into the beamline P10 of the synchrotron PETRA III at DESY~\cite{ESRF}. The well-collimated x-ray beam (wavelength $\lambda = 0.154$ nm) is deflected vertically to pass the layer of suspension in the shear-gradient direction~\cite{Denisov15}.  After loading, the samples are sealed with low-viscosity oil to prevent evaporation and guarantee sample stability over more than 4 hours, allowing us to measure samples repeatedly and reproducibly. Samples were initialized by a fixed protocol (preshear at $\dot{\gamma}=0.1$ s$^{-1}$ for 120 seconds, followed by 600 seconds rest). For each sample volume fraction, we apply oscillatory strain with frequency $\omega=\pi$, $2\pi$, $4\pi$, $10\pi$ and $20\pi~(rad/s)$ and amplitude $\gamma_0$ increasing from $\gamma_{0min} = 10^{-4}$ to $\gamma_{0max} = 0.4$ (forward cycle) and decreasing from $\gamma_{0max} = 0.4$ to $\gamma_{0min} = 10^{-4}$ (backward cycle). Each experiment consists of 100 points on a logarithmic scale, three oscillations averaged per cycle, leading to total duration of the experiment of around 1 hour for $\omega = \pi~rad/s$.

We use a Lambda detector to simultaneously monitor the scattered intensity from the sheared suspensions. The detector with pixel size $52 \times 52$ $\mu$m$^2$ and operating at a frame rate of 10 Hz, is placed at a distance of $D = 280$ cm, capturing wave vectors in the range \com{$q r_0= 0.5$ to $5$} in the shear direction--shear axis plane.

\section{Experimental Results and Discussion}
\begin{figure}
	\centering
	\subfigure
	{\includegraphics[width=0.8\columnwidth]{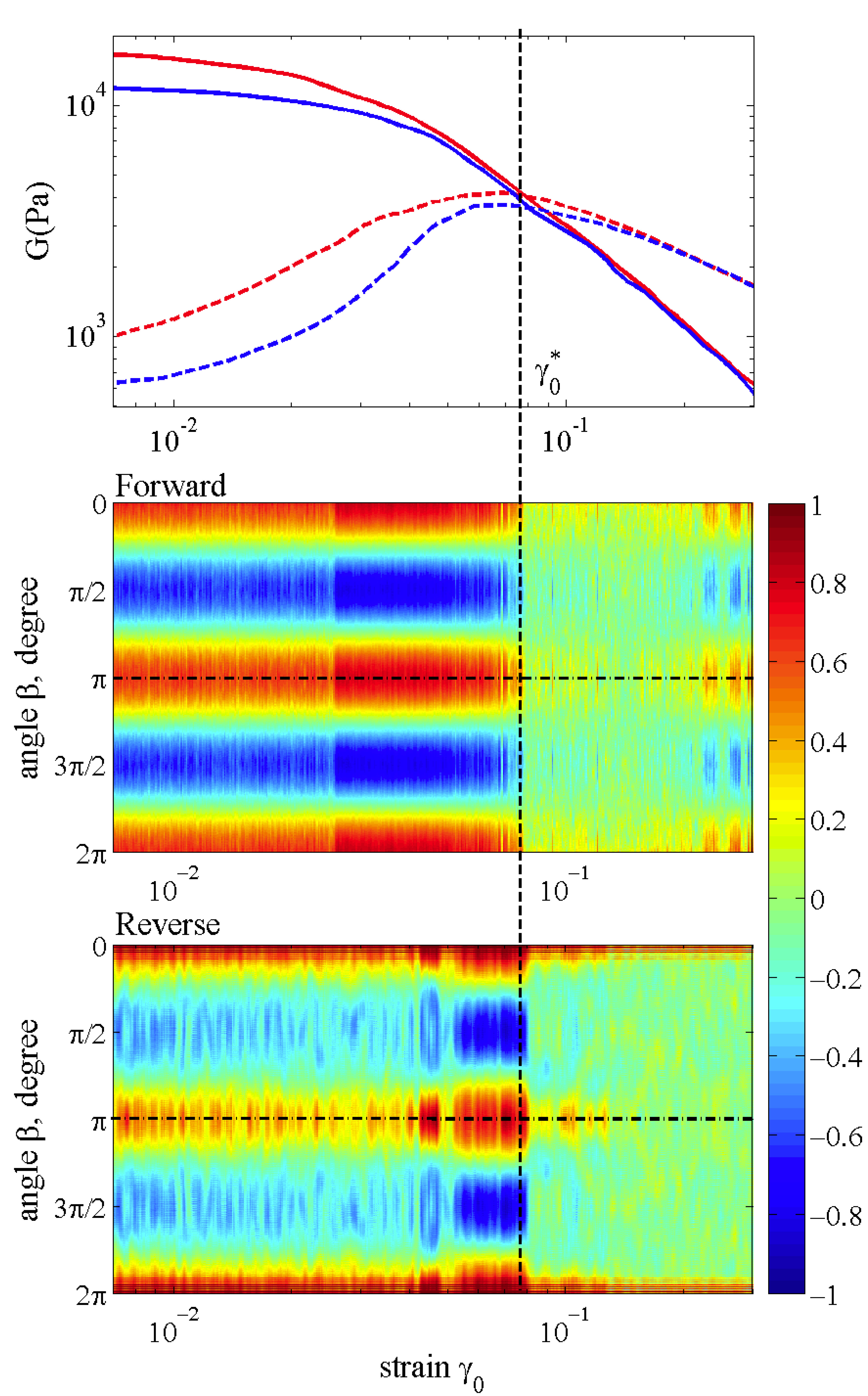}}
    \put(-196,210){(a)}
    \put(-196,110){(b)}
    \put(-196,5){(c)}
	\caption{(Color online) Sharp structural transition in the oscillatory rheology of a colloidal glass at $\phi=58\%$ and $\omega = 2\pi~(rad/s)$ (a) Elastic modulus $G^\prime$ (solid lines) and viscous modulus $G^{\prime\prime}$ (dashed lines) during increasing (red) and decreasing strain amplitude (blue). The intersection of $G^{\prime}$ and $G^{\prime\prime}$ in the forward cycle (increasing strain amplitude) is marked by $\gamma_0^*$. In the backward cycle (decreasing strain amplitude), this intersection occurs at slightly larger strain, indicating hysteresis. (b,c) Contour plots of the angular correlation of the scattered intensity indicate sharp symmetry change of the sheared structure. In the forward cycle (b, increasing strain amplitude), the two-fold symmetry vanishes abruptly at $\gamma_0^*$, indicating a sharp transition from elastic to viscous response. In the backward cycle (c, decreasing strain amplitude), the transition occurs at slightly higher strain, similar to the intersection of $G^\prime$ and $G^{\prime\prime}$ in (a).}
	\label{fig:angular}
\end{figure}

As we have shown in~\cite{Denisov15}, our setup combining x-ray scattering and rheology allows us to reveal a sharp symmetry change upon increasing strain amplitude: we observed a sharp transition from twofold symmetry characteristic of a sheared solid to isotropic symmetry characteristic of a slowly sheared liquid. This transition occurs at the intersection of the moduli, where the storage modulus, $G^\prime$, decreases below the loss modulus, $G^{\prime\prime}$, typical for the nonlinear rheology of soft glassy materials~\cite{Mason1997}, as shown by the red data in Fig.~\ref{fig:angular}a.
\begin{figure}
	\centering
	\subfigure
	{\includegraphics[width=0.8\columnwidth]{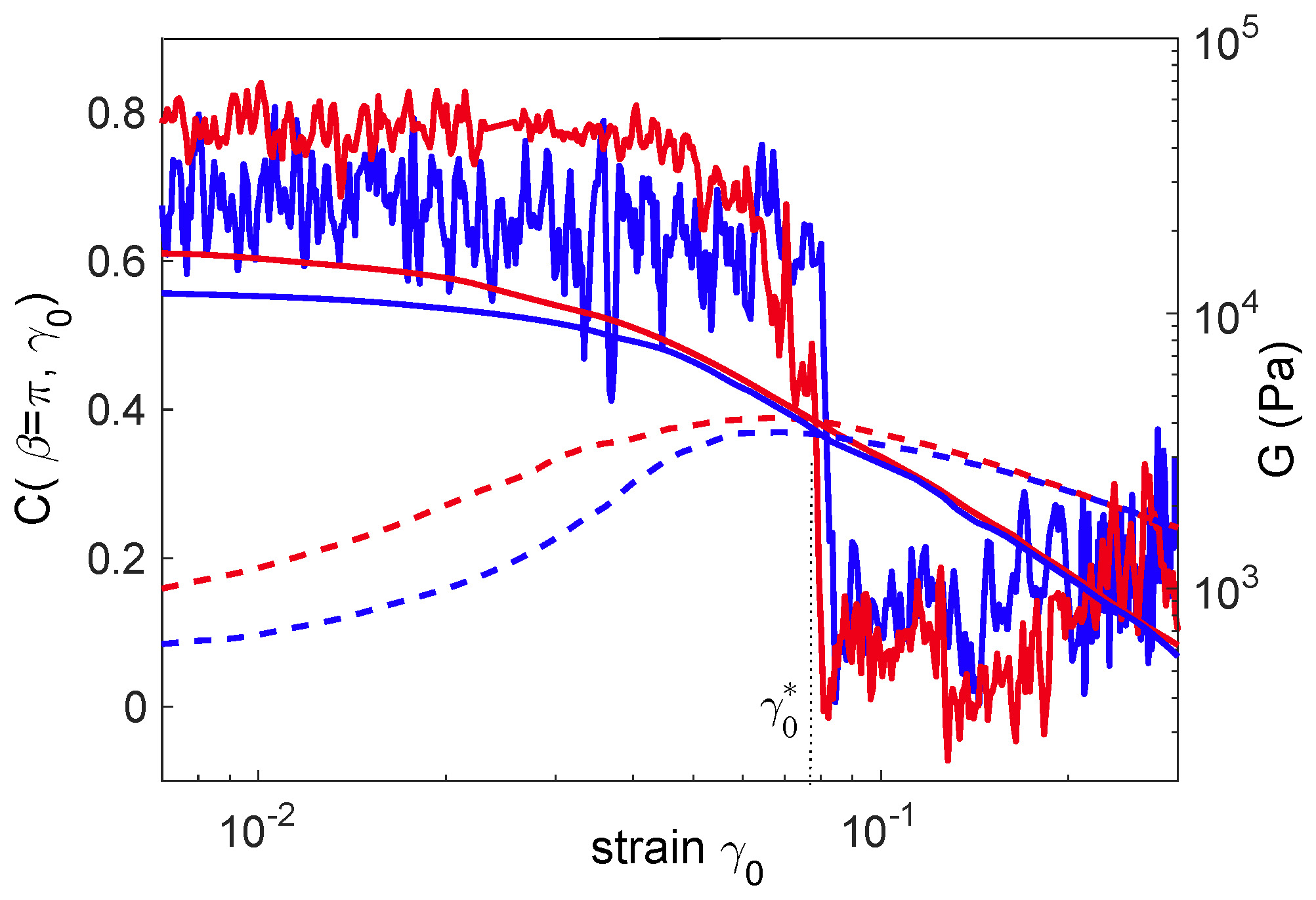}}
	\caption{(Color online) Reversible structural transition. The structural order parameter (left axis) characterizing the degree of two-fold symmetry from the peak value of the angular correlation function at $\beta=\pi$ is plotted as a function of strain amplitude together with the elastic and viscous moduli (right axis). Forward and backward shear are indicated by red and blue color, respectively. Sharp drop (red) and rise (blue) of the order parameter indicates the occurrence of a sharp, reversible structural transition at the intersection of $G^\prime$ and $G^{\prime\prime}$. }
	\label{fig:orderparam}
\end{figure}

\begin{figure}
	\includegraphics[width=0.49\textwidth]{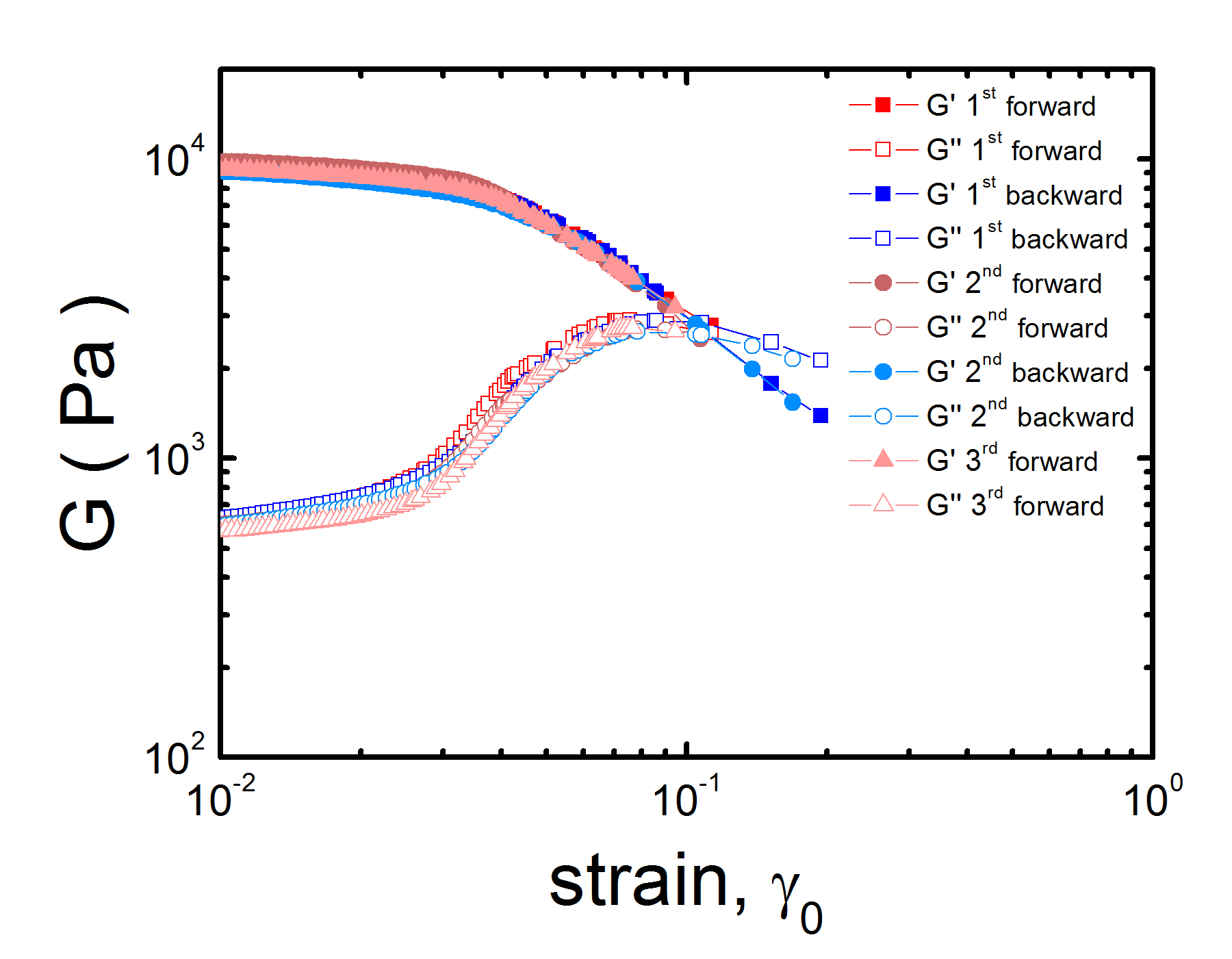}
	\caption{(Color online) Reproducibility of the rheological measurement. Elastic and viscous moduli, $G^\prime$ and $G^{\prime\prime}$ over several cycles of forward and backward strain sweep at frequency $\omega = 2\pi$ $~(rad/s)$ and volume fraction \com{$59\%$}. The overlap of multiple forward and backward cycles demonstrates excellent reproducibility. Only minor differences in the moduli (smaller than the symbol size) are observed between the runs.}
	\label{fig:repruducibility}
\end{figure}

\begin{figure*} [!]
	\includegraphics[width=0.55\textwidth]{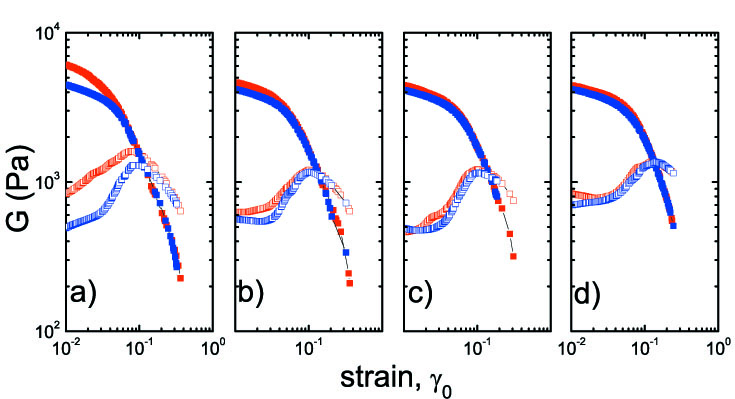}
    \includegraphics[width=0.37\textwidth]{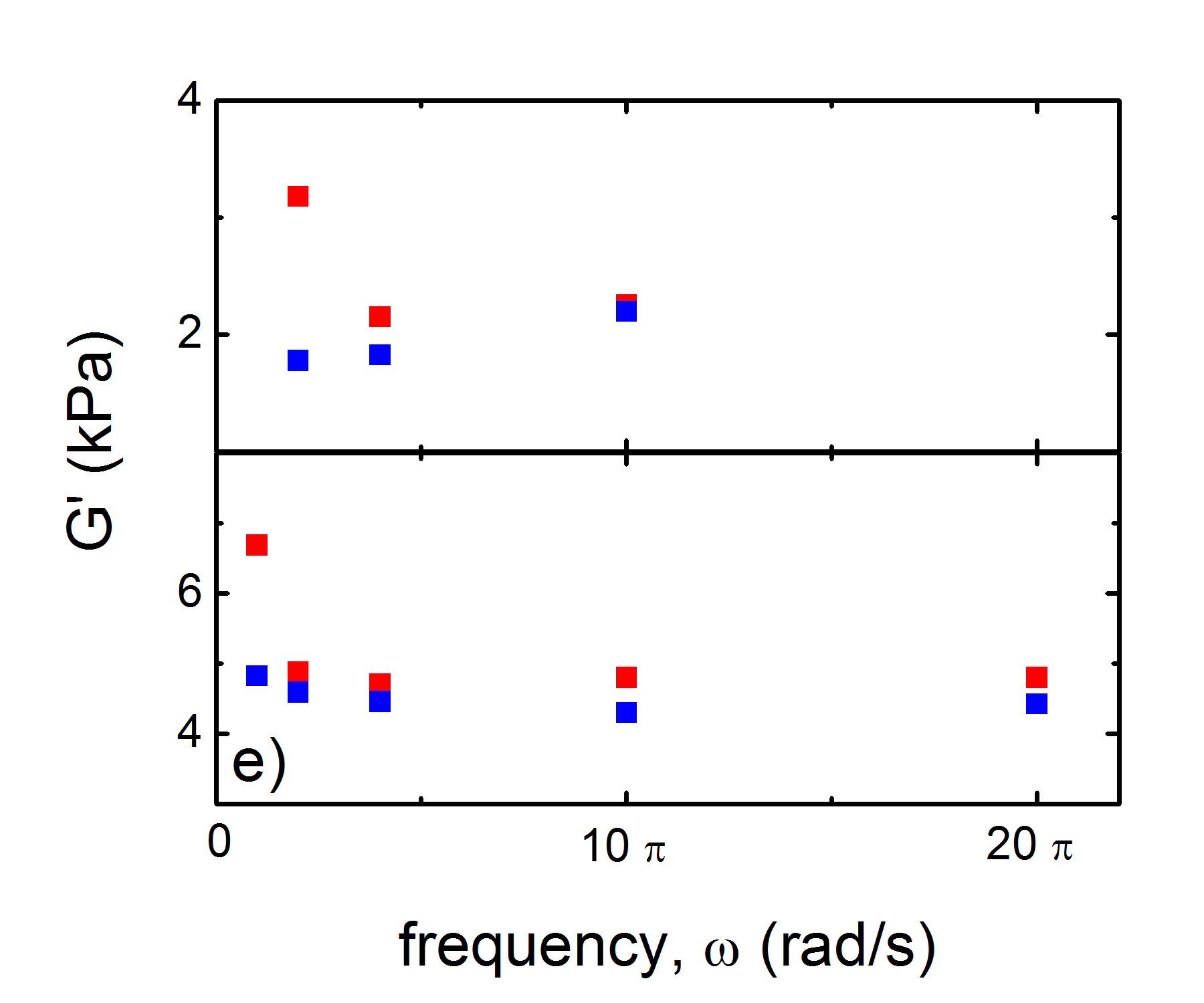}
	\caption{(Color online) Elastic and viscous moduli, $G^\prime$ and $G^{\prime\prime}$ of forward (red) and backward (blue) oscillatory shear at frequencies $\omega= \pi$ (fig. a), $2\pi$ (fig. b), $4\pi$ (fig. c), $20\pi$ $~(rad/s)$ (fig. d) at volume fraction $\phi=58\%$. (e) Plateau values of $G^\prime$ taken at $\gamma_0=10^{-4}$ as a function of frequency for forward (red) and backward (blue) shear at volume fractions $\phi=56\%$ (top) and $\phi=58\%$ (bottom). In both cases, the largest hysteresis is observed for the lowest frequency. Error bars are smaller than the symbol size.}
	\label{fig:rheo_vf58_2015}
\end{figure*}
The identified symmetry change shows up in all angle-dependent measures of the structure factor. \com{We focus on the first diffraction ring in the scattering plane, i.e. the nearest-neighbor peak. Its twofold symmetry at small strain indicates the elastic-like distortion of nearest-neighbor environments in the sheared glass~\cite{Denisov15}. This symmetry is most clearly demonstrated using angular correlations that pick out the underlying symmetry of the structure factor most efficiently (see appendix), as shown in Fig~\ref{fig:angular}b.
Twofold (p-wave) symmetry is clearly observed at small strains.} With increasing strain amplitude, the twofold pattern suddenly disappears at $\gamma_0 \sim 0.08$. This symmetry loss indicates the sudden transition to a liquid-like state. To demonstrate the sharpness of this transition, we follow the maximum of the angular correlation function located at $\beta = \pi$ \com{(dash-dotted horizontal line)}; this maximum quantifies the extend of two-fold symmetry and hence serves as order parameter of the transition: it is 1 for ideal two-fold symmetry and 0 for isotropic symmetry. As shown in Fig.~\ref{fig:orderparam} (red data), this order parameter drops indeed sharply to zero, manifesting the sudden loss of two-fold symmetry. Moreover, we find that the sudden symmetry loss occurs precisely at the intersection of the two moduli $G^{\prime}$ and $G^{\prime\prime}$, indicating that it is connected to the rheological loss of rigidity of the material. This sharp symmetry change reminds of first-order equilibrium transitions, but in the case here is induced by the applied oscillatory shear.

Exploiting the analogy with first-order transitions, the question is then whether this sharp symmetry change is reversible, and whether there is any hysteresis. To investigate the reversibility, we added a "backward" oscillatory shear cycle, i.e. we started from large strain amplitude in the nonlinear regime, and decreased the oscillatory strain amplitude down to small values in the linear regime. Remarkably, we find that indeed the transition reverses, as shown in Fig.~\ref{fig:angular}c. The two-fold symmetry re-appears spontaneously, and the order parameter jumps back to values of order 1, suggesting that the material abruptly acquires solid properties. We find that in the case here, both forward and backward transitions occur at very similar strain amplitude, and there is little hysteresis. Some hysteresis, however, occurs in the magnitude of the moduli $G^\prime$ and $G^{\prime\prime}$ and the magnitude of the order parameter. Both reflect the same trend: while the material clearly recovers its elastic properties during the backward transition as shown by both the order parameter and the moduli, the somewhat smaller magnitude of both compared to the forward cycle indicates that the material has not fully restored its initial rigidity. This gradual loss of rigidity is likely associated with restructuring of the material during the shear cycle, weakening it.

To analyse this hysteresis effect in more detail, we focus on the rheology and investigate its frequency dependence. We first confirm that the rheological measurements are well reproducible, as shown by several repetitions of forward and backward cycles superimposed on each other in Fig.~\ref{fig:repruducibility}. Forward cycles as well as reverse cycles closely overlap over several repetitions, demonstrating that the run-to-run variation of our rheological experiments are small (less than $10 \%$ of the magnitude of $G^\prime$ and $G^{\prime\prime}$). Second, when we repeat the experiment at varying frequency, large deviations between forward and reverse cycle eventually emerge at low frequency, as shown in Fig.~\ref{fig:rheo_vf58_2015}. Clearly, the growing difference between the forward (red) and backward (blue) cycle demonstrates that hysteresis effects grow with decreasing frequency.
%At the lowest frequency, The large difference between the forward and backward shear are observed, which almost completely vanish at the highest frequency.
This difference between forward and backward shear leads also to a shift of the crossing point of the moduli. In particular, the loss modulus $G^{\prime\prime}$ increases less steeply on the way back and exhibits a smaller maximum. This change of $G^{\prime\prime}$ is likely associated with restructuring processes occuring in the "liquified" state at large strain amplitude. Such processes, known as shear thinning in steady shear, typically reduce the viscosity to facilitate the flow. The higher slope of $G^{\prime\prime}$ in the forward cycle indeed suggests that the material exhibits stronger shear thinning than on the way back. In the backward cycle, the material then approaches the yield strain with a much more ordered structure and hence much lower viscosity than it had when it approached yielding coming from low strains. It follows that in the backward cycle, the suspension reaches yielding with lower values of $\eta$ and $G''$ compared to those in the forward route. Since $G'$, in contrast, is not much affected by the structural changes at large strain, the fact that $G''_{\leftarrow}<G''_{\rightarrow}$ while $G'_{\leftarrow} \sim G'_{\rightarrow}$, must necessarily imply that the intersection of $G^{\prime}$  and $G^{\prime\prime}$ shifts to larger strain. These results suggest that restructuring of the material leads to hysteresis effects at low frequency and long time scale, as is expected for slow relaxation. Similar frequency dependence is also observed for other volume fractions, both above and below the colloidal glass transition. This is demonstrated in Fig.~\ref{fig:rheo_vf58_2015}e, where we show the plateau values of $G^{\prime}$ for $\phi = 56\%$  (top) and $58\%$ (bottom). Hysteresis effects are always the largest at low frequency. \com{We note that while we confirmed that the hysteresis effects are reproducible over independent runs, the precise behavior over subsequent repetitive cycles needs further investigation.}

\begin{figure*} [!]
	\includegraphics[width=0.52\textwidth]{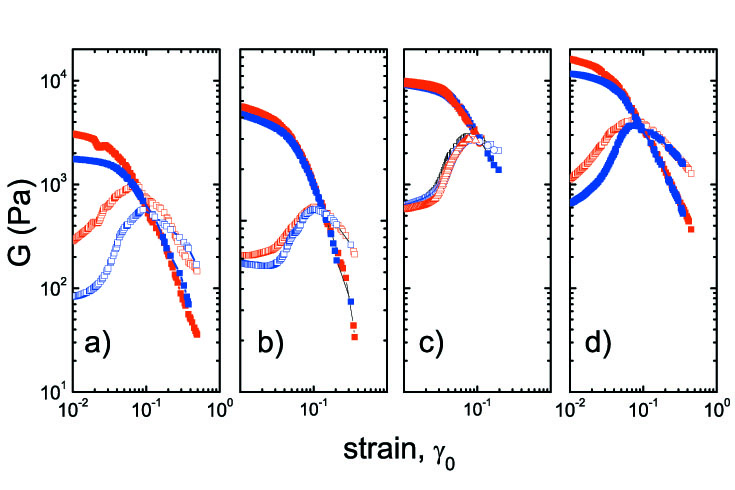}
    \includegraphics[width=0.4\textwidth]{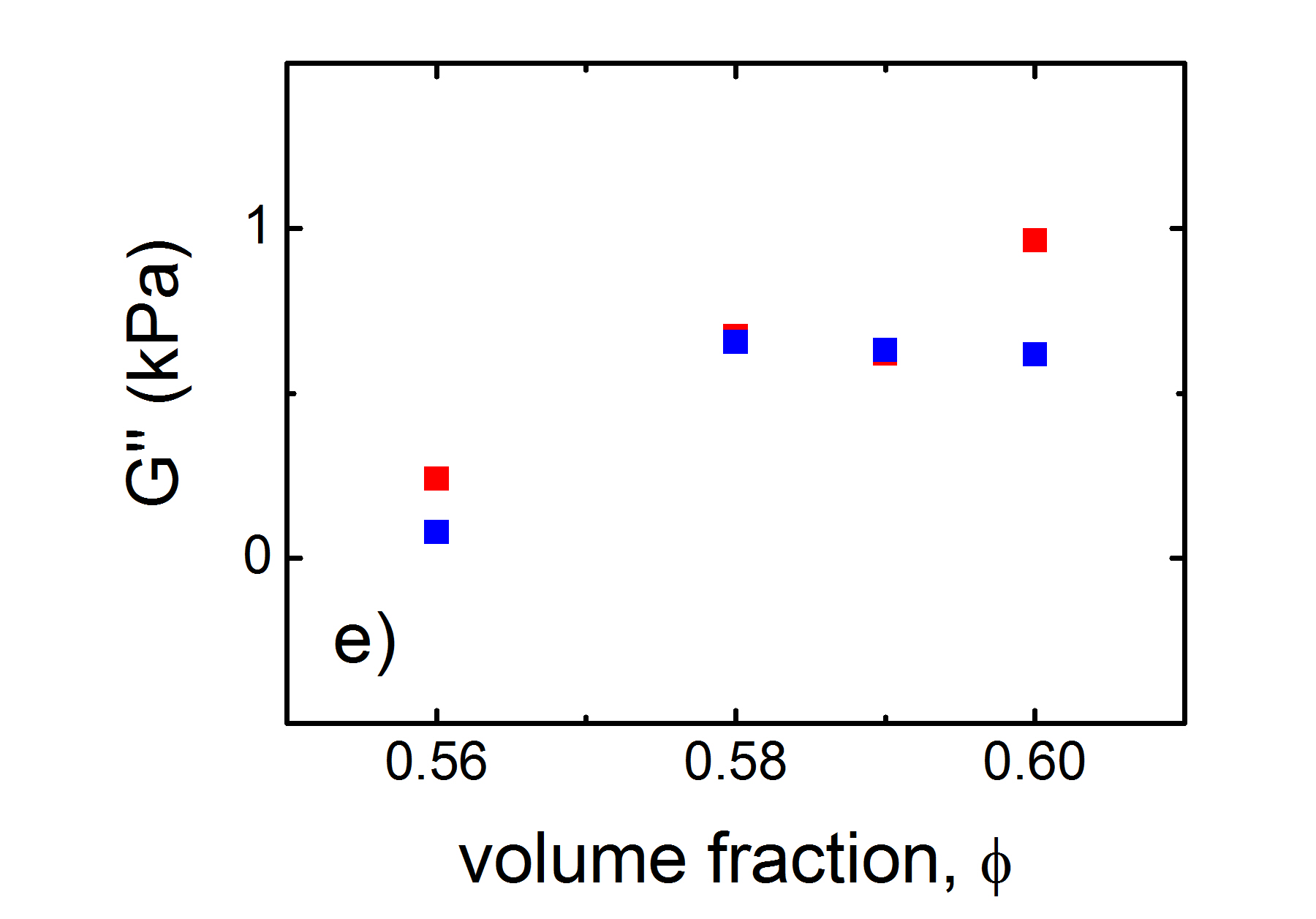}
	\caption{(Color online) Elastic and viscous moduli, $G^\prime$ and $G^{\prime\prime}$ of forward (red) and backward (blue) oscillatory shear at frequency $\omega = 2\pi$ $~(rad/s)$ for volume fractions $56\%$ (a), $58\%$ (b), $59\%$ (c), and $60\%$ (d). In panel (c), multiple forward and backward cycles are overlaid onto each other. The overlap demonstrates excellent reproducibility. (e) Plateau values of $G^{\prime\prime}$ taken at $\gamma_0=10^{-4}$ as a function of volume fraction $\phi$ at frequency $\omega = 2\pi$ $~(rad/s)$ of forward (red) and backward shear (blue). The error bars are smaller than the symbol size.}
	\label{fig:rheo_f1}
\end{figure*}

We also investigated the hysteresis as a function of volume fraction, and found a surprising non-monotonic relation. Rheological data for four different volume fractions $\phi=56\%$, $58\%$, $59\%$ and $60\%$ are shown in Fig.~\ref{fig:rheo_f1}. Interestingly, the hysteresis almost vanishes at around the glass transition (Fig.~\ref{fig:rheo_f1}b and c), while it becomes much more significant both below and above $\phi_g=58\%$. At the glass transition, the rheology is closely reversible: the moduli of the forward and backward cycle overlap nicely, even over several consecutive cycles. In contrast, significant hysteresis emerges for densities deep inside the glass (Fig.~\ref{fig:rheo_f1}d), and even more so below the glass transition (Fig.~\ref{fig:rheo_f1}a). The fact that the strongest hysteresis occurs for the lowest investigated volume fraction ($\phi=56\%$) lends credence to our interpretation that the hysteresis effect is due to shear-thinning processes. At this volume fraction, the system is in a viscous, supercooled liquid state. Since it is relatively more dilute compared to the glass, the particles can restructure more easily, a process that lowers the viscosity due to the well-known shear-thinning effect upon increasing the strain rate. This is reflected in a much steeper power-law decrease of $G''\propto\eta\sim\gamma_{0}^{-0.7}$, compared to the glass at higher $\phi$ where $G''\propto\eta\sim\gamma_{0}^{-0.5}$. On the other hand, in the supercooled state the lower slope of $G^{\prime\prime}$ after shear reversal directly indicates the lower viscosity due to restructuring. We give an overview of the hysteresis behavior by plotting the plateau value of $G^{\prime\prime}$ as a function of volume fraction $\phi$ in Fig.~\ref{fig:rheo_f1}e.

\begin{figure}
	\centering
	\subfigure
	{\includegraphics[width=0.9\columnwidth]{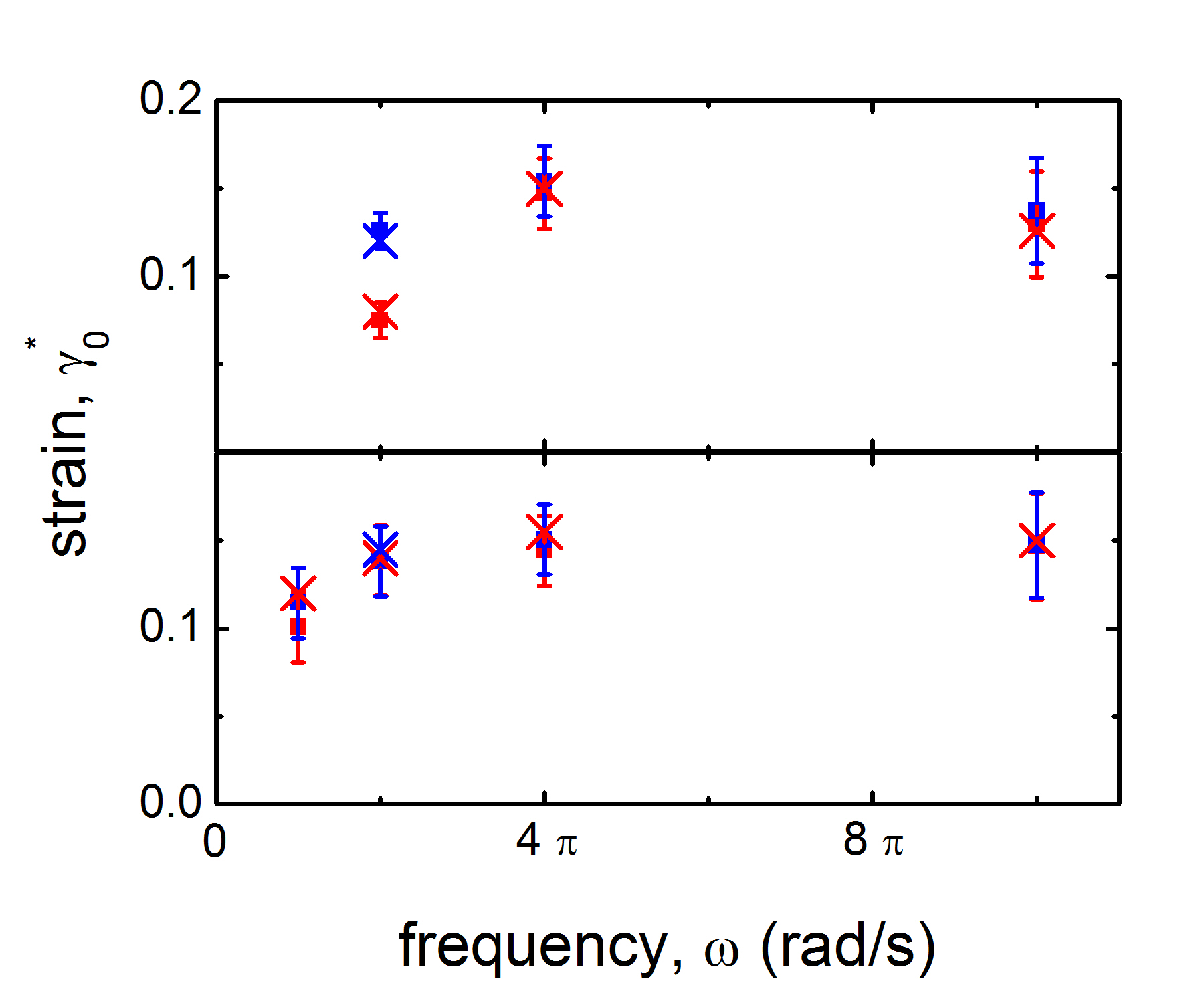}}
	\caption{(Color online) Strain amplitudes at yielding as a function of frequency for forward (red) and backward shear (blue) at volume fractions $\phi=56\%$ (top) and $\phi=58\%$ (bottom). Squares and error bars indicate the intersection of $G^\prime$ and $G^{\prime\prime}$, while crosses indicate the structural transition. Both coincide within error bars. A few structural transition data points are missing due to beam loss.}
	\label{fig:yielding_phi}
\end{figure}

How does this rheological hysteresis relate to the structural transition? Are similar hysteresis effects observed in the sharp symmetry change? To address this, we investigated in detail the structural transition as a function of frequency and volume fraction. We first confirm that for all frequencies and volume fractions within the range investigated, the sharp structural transition is a robust feature, and always occurs both in the forward and backward cycle. Remarkably, when we determine the location of the transition $\gamma^*$, we find that it always coincides with the intersection of $G^\prime$ and $G^{\prime\prime}$. This is shown in Fig.~\ref{fig:yielding_phi}, where we plot the location of the structural transition together with that of the intersection of the moduli. Whenever the intersection of $G^\prime$ and $G^{\prime\prime}$ shifts due to hysteresis, the structural transition shifts as well so that they always coincide, as demonstrated by the overlay of the symbols in Fig.~\ref{fig:yielding_phi}. This means that the coincidence of the structural transition and the intersection of the moduli is a robust experimental feature that remains valid even when the transition shifts due to hysteresis. This is most evident for $\phi=56\%$ at low frequency, where there is a large mismatch between $\gamma_0^*$ in the forward and backward cycle.
%We also note that the largest value of yield strain is observed for the sample right at the glass transition, $\phi=58\%$. This fact implies that the most ductile glass, which can sustain the largest strain prior to mechanical failure, is the one prepared right at the glass transition. Denser glasses are comparatively more brittle.
We have shown in a previous publication \cite{Denisov15} that the intersection point where $G^\prime = G^{\prime\prime}$ is equivalent to the equality of microscopic affine and non-affine components underlying the macroscopic deformation. Therefore, the robust coincidence of structural transition and intersection of the moduli indicates that it is the balance of affine and non-affine components of deformation that governs the rheological solid-liquid transition of the material. This robust principle motivates a simple theoretical model of non-affine deformation underlying the observed trends.

\section{Mean-field model of yielding}

\subsection{Nonaffine model for $G^\prime$}

We model the strain-dependent weakening of the material with a microscopic model of affine and non-affine deformation. In this model, the decrease of $G^\prime$ with increasing strain amplitude is due to shear-induced loss of connectivity. Upon application of shear, nearest neighbors of any tagged particle tend to be removed in the extensional sectors, whereas almost no new neighbors move in along the compression sectors due to excluded volume~\cite{Zausch09,Chikkadi12}. The resulting net loss of mechanically active nearest-neighbours leads to connectivity loss and a weakening of the structure with increasing strain~\cite{Zaccone11}. This effect has been confirmed  experimentally in the pair correlation function determined with confocal microscopy during start-up shear deformation~\cite{Dang2015}. The lost neighbours migrate into the free volume pockets as there is a favourable chemical potential gradient in that direction, which is reflected in the re-distribution of free volume. The loss of nearest neighbors and concomitant loss of rigidity leads to increasing non-affine displacements until, at yielding, the effective number of nearest neighbors is barely enough to sustain the nonaffine displacements required to keep mechanical equilibrium~\cite{Zaccone13b}.
%while no energy is left to react to the deformation.
This point defines the transition from solid to liquid at which the material starts to flow.

\begin{figure}
	\includegraphics[width=0.5\columnwidth]{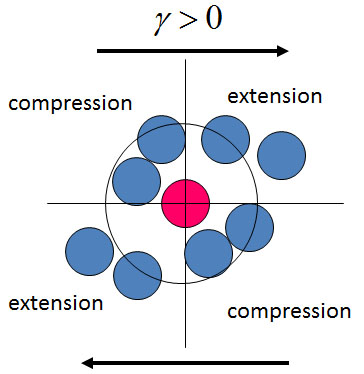}	
	\caption{(Color online) Schematic of nearest neighbor loss under applied shear. Upon application of shear, particles move out of the cage in the extension direction, and move in along the compression direction. Because of the hard-core potential, particles move in much less than they move out, leading to net loss of connectivity.}
	\label{fig:shearmodel}
\end{figure}

To model the resulting decrease of the elastic modulus, we first consider the affine part of the shear modulus $G_A=\frac{1}{5\pi}\frac{\kappa\phi}{\sigma}n_{b}$ in the linear regime, according to the Born-Huang theory of lattice dynamics~\cite{Born1954}. Here, $n_b$ is the number of nearest neighbors and $\kappa$ the spring constant associated with a nearest-neighbor bond. In our hard-sphere glass, "bonded" neighbors arise from the entropic attraction (akin to depletion attraction): basic statistical mechanics~\cite{Hansen2005}, relates the first peak of the radial distribution function $g(r)$ to an attractive minimum in the pair potential of mean force $V_{\mathrm{eff}}/kT=-\ln g(r)$. This also defines the elastic spring constant as
$\kappa=[d^{2}V_{\mathrm{eff}}/dr^{2}]_{r=\sigma}$ between two bonded neighbors. The number of bonded neighbors is given by the integral of the first peak of $g(r)$, which yields $n_{b}^{0} \approx 12$ for the static hard-sphere glass. The situation changes under applied shear, as shown schematically in Fig.~\ref{fig:shearmodel}: Particles become crowded in the compression sector of the shear plane, whereas they become dilated in the extension sector. Because of the strong excluded volume interactions (the nearest neighbors cannot come closer to a selected particle than its excluded volume), the particle increase in the compression sector cannot balance the particle loss in the extension sector, leading to a net loss of particles. This is indeed what we observed experimentally when we resolved the pair correlation function along the extension and compression direction~\cite{Dang2015}: In the extension direction, the first maximum of the pair distribution function $g(r)_{max}$ decreases continuously, while in the compression direction, this number increases only slightly to saturation.

\begin{figure}
	\includegraphics[width=0.9\columnwidth]{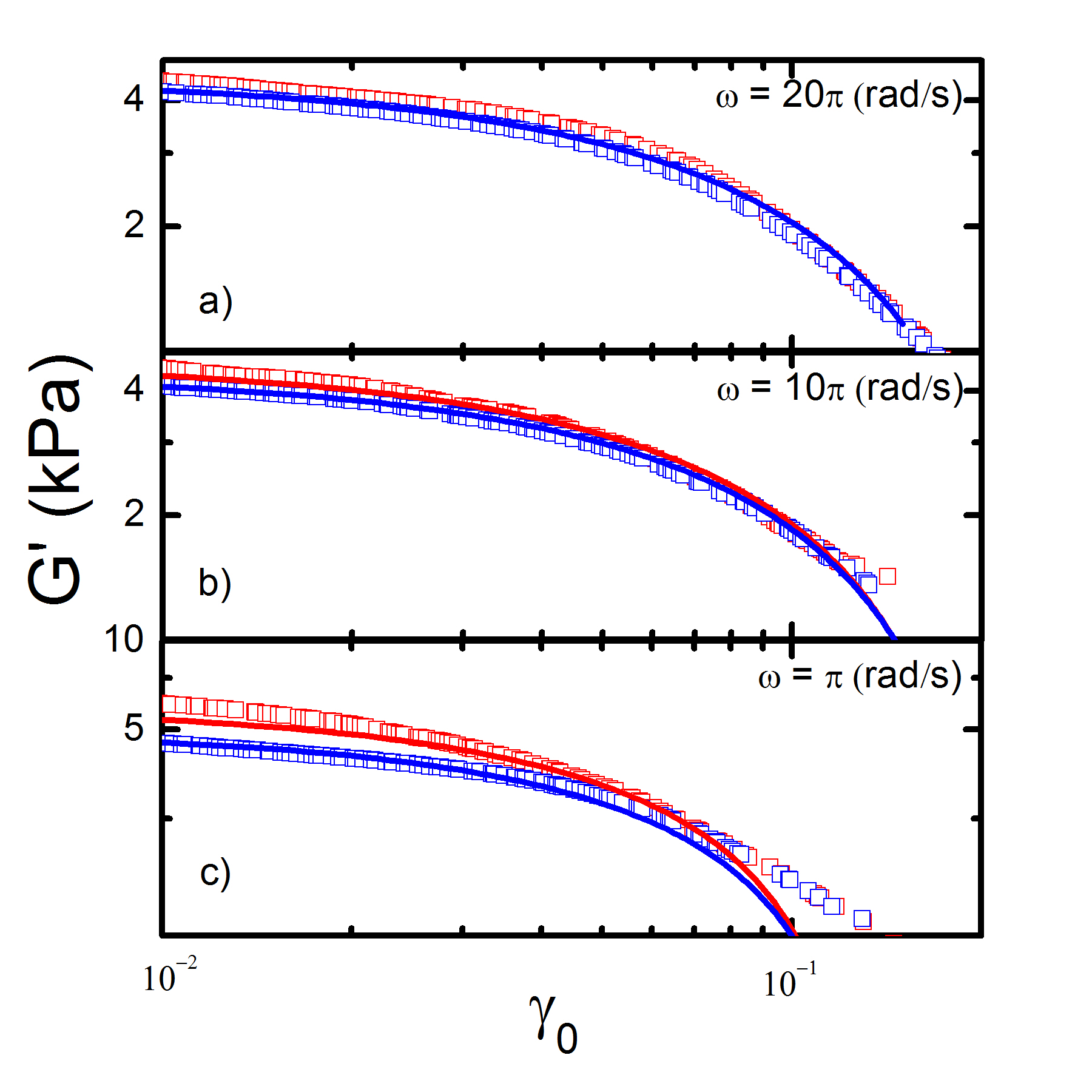}
	\caption{(Color online) Storage modulus $G^{\prime}$ fitted by the non-affine model, in forward and backward shear for $\phi=58\%$ and $\omega=20\pi$ (a), $\omega=10\pi$ (b), and $\omega = 1\pi$ $~(rad/s)$ (c). For the highest frequency (a), the agreement is perfect because shear thinning is negligible, i.e. the high deformation rate ensures that no relaxation occurs. Thus, the number of nearest neighbors remains unchanged, $n_{b,0} = n_{b,end} = 12$.  Increasing deviation occurs at lower frequency, indicating increasing relaxation. Accordingly, the number of neighbors at the end of the cycle $n_{b,end} = 11.5$ for both $\omega = 10\pi$ and $1\pi$ $~(rad/s)$, while the amplitudes $A_f=A_b=-4.6$ for $\omega=10\pi$ $~(rad/s)$ (b) and $A_f=-4.9$ and $A_b=-4.6$ for $\omega=1\pi$ $~(rad/s)$ (c).}
	\label{fig:fitting}
\end{figure}

Assuming that the local cage dynamics is governed by the Smoluchowski equation with shear~\cite{Dhont}, \com{we found previously ~\cite{Dang2015} that the number of nearest neighbors (proportional to the first peak of $g(r)$) decreases exponentially with strain, $n_b(\gamma) = n_{b}^{0}\exp(-A\gamma)$ corresponding to a decrease of the first-peak of $g(r)$ in the extension direction according to $g_c(r) = 3.07\exp(-A\gamma)$}. The numerical factor $A$ in the exponential decay follows from the fit to the experimental data for $g(r)$ presented in \cite{Dang2015}. Since this parameter represents the extent of the shear-induced microscopic connectivity loss, its fitted values, as discussed below, may vary depending on the shear protocol, the glass volume fraction, and the frequency.

As a result of the reduced connectivity, there are increasing non-affine contributions to the shear modulus, as shown recently also in numerical simulations~\cite{Priezjev}. According to Alexander~\cite{Alexander98}, the nonlinear (non-affine) contribution to the shear modulus can be written as resulting from a Taylor expansion in the free energy up to third order in $\gamma$, which gives a first-order in $\gamma$ correction to the modulus as:
$G_{NA}= \frac{1}{5\pi}\frac{\kappa\phi}{\sigma}~(n_{b}^{c}+C\gamma)$, where $C=3$ is a phenomenological constant from the non-affine free energy expansion and the critical coordination number $n_{b}^{c}=6$~\cite{Zaccone11} for central-force interactions in the quasistatic limit. The parameter $C$ thus contains the information about how sensitively the nonaffine contributions depend on the strain amplitude. The total modulus is given by the sum of affine and non-affine contributions, which, because non-affine parts contribute negatively (against the internal force field to restore force balance), reads: $G = G_{A}-G_{NA}= \frac{1}{5\pi}\frac{\kappa\phi}{\sigma} [n_b(\gamma) - n_{b}^{c} - C\gamma)]$.

In the low-strain limit, $\gamma\rightarrow 0$, the nonaffine contribution reduces to $G_{NA}= \frac{1}{5\pi}\frac{\kappa\phi}{\sigma}6$, which, combined with the affine contribution leads to the scaling $G \sim (n_b - 6)$, valid at zero frequency for athermal disordered solids. At high frequency $\omega \gg 1/\tau$, it is widely accepted that the particles move in a dominantly affine way in the elastic regime before yielding~\cite{Hansen2005}, and $G \simeq G_{A}$ is a good approximation. In physical terms, at high frequency the particles have not enough time to relax into their nonaffine positions and are stuck in the higher-energy affine positions as the deformation is quickly reversed. This implies that the critical coordination number $n_{b}^{c}$ associated with the nonaffine relaxation, effectively must decrease all the way from $n_{b}^{c}=6$ at $\omega=0$ to $n_{b}^{c}=0$ at $\omega=\infty$, in order to recover $G =G_{A}$ as the infinite-frequency modulus. These limits are fully consistent with the well-documented existence\cite{Hansen2005,Larson} of two limiting plateaus in the frequency-dependent modulus of glasses, namely a low-frequency plateau where the modulus is the lowest in value and the nonaffinity is the highest, and a high-frequency plateau where the modulus is affine, in good approximation.  Hence, the critical connectivity parameter $n_{b}^{c}$ which controls the extent of nonaffinity, is expected to decrease with increasing $\omega$ and we adjust it as a function of frequency to fit the oscillatory shear data of $G'$ at varying frequencies. The shear-induced connectivity decay coefficient $A$ will be shown to not depend much on the frequency and the constant $C=3$ remains unchanged in the fitting process.

We independently fit the forward and backward shear cycle. We use $n_{b}^{0}$ to indicate the number of nearest neighbors before starting the shear, which for dense liquids and glasses of spherical particles is $n_{b}^{0}=12$. We use $n_{b}^{end}$ to define the number of neighbours at the end of the backward shear cycle, which may be different from $n_{b}^{0}$ due to shear-induced restructuring: some particles may have irreversibly migrated into "free volume" pockets where they effectively behave similarly to rattlers in packings, i.e. without exhibiting any mechanically active contact with other particles. This mechanism implies that irreversible motions of this kind should ultimately deplete, on average, the cage of neighbours at the end of the cycle. Since $n_{b}^{end}$ is a fitting parameter, whenever we will find $n_{b}^{end}<n_{b}^{0} = 12$, this is a strong indication that the deformation has a finite degree of irreversibility, reflected in a lower $G'$ in the backward cycle.

The final formula used for the fitting of $G'$ is thus
\begin{equation}
G'=K\{{n_{b}(\gamma)-[n_{b}^{c}(\omega)+C\gamma]}\}
\label{eq_G}
\end{equation}
where $n_b(\gamma) = n_{b}^{0}\exp(-A\gamma)$ for the forward cycle, and $n_b(\gamma) = n_{b}^{end}\exp(-A\gamma)$ for the backward one, while $n_{b}^{c}(\omega)$ is fitted within a reasonable range and varies (decreases) with $\omega$. $K$ is a fitting prefactor which is proportional to $\kappa \phi/\sigma$.

\subsection{Effect of frequency on $G^\prime$}
To check this model, we fit $G^\prime$ in eq.~\ref{eq_G} to the measured elastic modulus from Fig.~\ref{fig:rheo_vf58_2015}. As shown in Fig.~\ref{fig:fitting}, the model captures the gradual decrease of $G^\prime$ all the way from the linear regime at small strain to the nonlinear regime. Perfect agreement is obtained for the highest frequency: both forward and backward shear curves overlap, and there is one and the same theoretical curve describing both. In particular,  $n_{b}^{end}=n_{b}^{0} = 12$, indicating that no structural change has occurred. The situation changes at lower frequency, with small deviations at $10\pi$ $~(rad/s)$ (Fig.~\ref{fig:fitting}b), and larger ones at $\pi$ $~(rad/s)$ (Fig.~\ref{fig:fitting}c). The $G^\prime$ curves of the forward and backward shear cycle become different and the fit is good only within the pre-yielding regime, i.e. for $\gamma_0<0.1$. In fact, deviations occur in the limit of high strain, where the experiments show a decreasing power-law trend of both $G^\prime$ and $G^{\prime\prime}$, indicating the presence of shear-thinning phenomena.
%where the resistance to flow is reduced through the shear-induced ordering of particles in the fluid-like suspension.
Shear thinning is not included in our simple model for $G'$ and therefore the power-law dependence is not captured; \com{the resulting restructuring reduces the non-affinity, causing the decrease of $G^\prime$ to be smaller than predicted by the non-affine model, which explains the discrepancy in the post-yielding regime.} At high frequency, this shear-thinning is less prominent because the short time scale does not allow the particles to restructure into the layered shear-thinning configuration.

Correspondingly, the parameter $n_{b}^{end}$, a sensitive measure of irreversibility in the model, becomes smaller than $n_{b}^{0}$ at these lower frequencies. From the fits of the data we indeed observe that $n_{b}^{end}=11.5 < 12$, implying a finite extent of irreversible motions, at the lowest frequency $\omega=\pi~~(rad/s)$. This means that, on average $0.5$ particles in the original quiescent glassy cage has been lost, and migrated irreversibly into free volume regions. The $G'$ elastic plateau at the end of the cycle is thus lower than the $G'$ elastic plateau at the start of the deformation. With increasing frequency, the values of $n_{b}^{end}$ tend to increase until ultimately for the highest frequency investigated here, $\omega=20\pi$ $~(rad/s)$, we find $n_{b}^{end}=12$, implying a fully reversible deformation cycle and also $G'$ having the same plateau value for both the forward and backward regimes.

Finally, it is instructive to look at the fitted values of the nonaffinity parameter $n_{b}^{c}$. We find that indeed $n_{b}^{c}=6$ at the lowest frequencies of $\pi~~(rad/s)$ and $10\pi~~(rad/s)$, implying that we are still close to the low-frequency non-affine plateau of the shear modulus. With increasing frequency, the value of $n_{b}^{c}$ decreases until at $\omega=20\pi~~(rad/s)$ we find $n_{b}^{c}=3$, closer to the affine high-frequency plateau, where $n_{b}^{c}$ vanishes.

\subsection{Nonaffinity-based interpretation of $G^{\prime\prime}$}

To better understand the behaviour of the viscous modulus, $G^{\prime\prime}$, we now focus on dissipative processes and their description within the non-affine response theory. Our treatment will allow us to make qualitative predictions about the behaviour of hysteresis and dissipation as a function of frequency and volume fraction. We here present only the essential results of the theory and derive its consequences for the viscous modulus, with particular emphasis on the hysteresis. The full nonaffine model of viscous dissipation, derived here for the first time, is presented in appendix B.

Similar to the elastic modulus described above, the dissipative function contains two contributions: one is the standard contribution related to viscosity and the loss modulus $G''$, while the second contribution discussed here for the first time is associated with the time-dependence of nonaffine displacements in the amorphous solid. While this latter, non-standard dissipation does not directly contribute to the viscosity (which is defined based on affine displacements~\cite{Landau}), it does alter the total energy of the system and may explain the decrease in the plateau of the shear modulus at the end of the deformation cycle in the limit of high packing fraction.

Recall that the experimental data show pronounced hysteresis both at the lowest and the highest volume fraction, while for $\phi \sim \phi_g$, no hysteresis occurs. The hysteresis, i.e. the fact that the plateaus of $G'$ and $G''$ are lower at the end of the deformation cycle, reflects the fact that a quote of internal energy of the solid has been lost to dissipation during the cycle. We define the dissipated energy per time $\dot{E}=-2\Psi_{tot}<0$ during the shearing, where the Rayleigh dissipation function $\Psi_{tot}>0$. Hence, the hysteresis of $G^\prime$ and $G^{\prime\prime}$ is directly related to the behaviour of the dissipation function $\Psi_{tot}$.

As shown in Appendix B, the total dissipation function $\Psi_{tot}$ in the solid regime can be written as
\begin{equation}
\Psi_{tot}=\frac{1}{2}\eta\dot{\gamma_{0}}^2+B\mid\gamma_{0}\mid\dot{\gamma_{0}},
\label{eq2}
\end{equation}
where the first term is the usual (affine) viscous dissipation in liquids and solids~\cite{Landau} controlled by the viscosity $\eta \sim G''$ at $\omega\rightarrow0$, and the second term arises from the finite rate of nonaffine motions in the amorphous solid. The second, non-standard contribution, is derived in Appendix B. Here we defined $\dot{\gamma_{0}}=\gamma_{0}\omega$, and the prefactor  $B$, which is independent of $\gamma_{0}$ and $\dot{\gamma_{0}}$, relates to the entropy change of the system, $B\sim(\partial S/\partial E)^{-1}$, where $E$ is the internal energy of the system, i.e. the sum of potential and kinetic energy. The above relation can be justified on the basis of the following argument: Working with the particle nonaffine displacement $x_{i}$, the dissipative function is defined in terms of the entropy production rate $\dot{S}$ by the relation: $2\Psi_{NA}=\dot{S}=(dS/d\underline{x}_{i})\cdot (d\underline{x}_{i}/dt)$. Upon introducing the Hessian (dynamical matrix) $\underline{\underline{H}}_{ij}$, and the thermodynamic relation $(\partial S/\partial E)^{-1}dS=-dR_{\mathrm{min}}$, where $dR_{min}$ is the reversible work equivalent to the work done by the nonaffine displacement $x_{i}$, this gives $\Psi_{NA}=\frac{1}{2(\partial S/\partial E)^{-1}}\underline{\underline{H}}_{ij}\underline{x}_{j} \dot{\underline{x}}_{i}\sim C \mid\gamma_{0}\mid\dot{\gamma_{0}}$.

In a hard-sphere colloidal glass, the main contribution to $E$ comes from the kinetic energy of the particles. Upon increasing $\phi$, the free volume decreases, and so does the kinetic energy of the particles, along with the decrease of mean square velocity fluctuations. Eventually, the kinetic energy becomes zero at random close packing, and $B\sim(\partial S/\partial E)^{-1}\rightarrow0$. Hence, $B$ in equation~(\ref{eq2}), is a decreasing function of $\phi$.
If we look at the viscous term, instead, we know that the viscosity of hard-sphere suspensions is an increasing function of $\phi$, and it eventually diverges at random close packing, with a power-law divergence. Hence, the two terms in equation~(\ref{eq2}) have opposite qualitative trends as a function of $\phi$.

The hysteresis observed in the experiments should be directly related to the dissipated energy: because any hysteresis implies changes of the structure, the larger the dissipated energy, the larger the restructuring and hence the larger the hysteresis. In an attempt to explain the nonmonotonic hysteresis trend with $\phi$, we can hence speculate that in the range $\phi=56\%-58\%$, the decrease of $B\sim(\partial S/\partial E)^{-1}$ with $\phi$ is stronger than the increasing trend of $\eta$. This is meaningful if one recalls that \com{experimental data suggests~\cite{Weeks12} the viscosity starts to vary strongly with $\phi$ only in the vicinity of random close packing~\cite{Larson-book,vanderVaart2013}, whereas the kinetic energy, which controls the decrease of $B$, has a very strong decrease upon entering the glass regime at $\phi=58\%$}. Furthermore, if the driving is slow, the viscous contribution, the first term in the above equation, can be considered small in the range $\phi=56\%-58\%$. Hence the decrease in hysteresis and dissipation upon going from $\phi=56\%$ to $\phi=58\%$ can be attributed to the fact that the term $\frac{1}{2}\eta\dot{\gamma_{0}}^2$ remains relatively small in this regime, whereas the term $B\mid\gamma_{0}\mid\dot{\gamma_{0}}$ decreases significantly upon increasing $\phi$. The minimum dissipation is achieved around $\phi=58\%$ where no hysteresis is observed. Upon further increasing $\phi$, however, the viscosity starts to increase more abruptly than the kinetic energy decreases, and the dissipation starts to increase as well. This results in an increase in hysteresis observed at $\phi=60\%$. We can hence understand the hysteresis trend qualitatively based on the non-affine model of dissipation in glasses.

\section{Conclusion}
We have demonstrated reversibility and hysteresis of the sharp structural symmetry change in the oscillatory shear of colloidal glasses. Upon increasing strain amplitudes from the linear into the nonlinear regime, the glass exhibits a sharp, reversible transition from an elastic to a liquid-like response, indicated by the sharp loss of shear-induced structural anisotropy active in the solid (which has a non-negligible affine deformation component). Upon varying frequency and volume fraction, this sharp transition always occurs, and always coincides with the intersection of the rheological moduli $G^\prime$ and $G^{\prime\prime}$. Looking in detail at the elastic and viscous moduli, we found that hysteresis effects arise at low frequency due to restructuring of the glass at these low rates and long time scales. We model this hysteresis behavior with a non-affine model that accounts for shear-induced loss of connectivity and rigidity of the glass. Using this model, we can describe the strain dependence of the elastic modulus quantitatively all the way from the linear to the nonlinear regime. Hysteresis occurs due to the additional shear thinning in the liquid state of the suspension after yielding, leading to restructuring. The corresponding dissipated energy arises as a competition between the regular dissipation in liquids and the entropy change associated with restructuring due to time-dependent non-affine particle motion.

\section{Acknowledgements}
The authors thank P. Lettinga for useful discussions. We thank DESY, Petra III, for access to the x-ray beam, proposal I-20130124 EC. We thank ESRF, DUBBLE beamline and NWO for for access to the x-ray beam, proposal 26-02-715.
This work was supported by the Foundation for Fundamental Research on Matter (FOM) which is subsidized by the Netherlands Organisation for Scientific Research (NWO).

\section*{Appendix A: Angular correlation function}
From the recorded diffracted intensity, we determine the structure factor $S(\textbf{q})$ by subtracting the solvent background and dividing by the particle form factor determined from dilute suspensions. In the linear elastic regime (low strain amplitude), we observe a characteristic p-wave distortion that is consistent with an elastic distortion of the nearest-neighbor structure due to the applied shear. To bring out this underlying symmetry most clearly, we focus on the first diffraction ring $S_1(\alpha)$ and compute angular correlations of the fluctuations of $S_1$ using
\begin{multline}
C(\beta)= \\
\frac{\int_0^{2\pi}(S_1(\alpha+\beta)-<S_1(\alpha)>)(S_1(\alpha)-<S_1(\alpha)>)d\alpha}
{\int_0^{2\pi}(S_1(\alpha)-<S_1(\alpha)>)^2 d\alpha}.
\label{eq1}
\end{multline}
Here,  $\alpha$ and $\beta$ are polar angles in the diffraction plane, and we integrate over the angle $\alpha$ as a function of the correlation angle $\beta$. Possible effects of elliptical distortion of the first ring are reduced by averaging radially over a range of wave vectors ($\Delta q \sim2w_1$) around $q_1$, where $w_1$ is the width of the nearest neighbor peak. We define the peak value $C(\beta = \pi)$ of the angular correlation function as structural order parameter; this allows us to measure the symmetry change as a function of applied strain.

\section*{Appendix B: Nonaffine contribution to dissipation (Eq.2)}
The elastic deformation of amorphous solids can be described and understood within the Born-Huang expansion of deformation free energy of solids, provided that the role of structural disorder is properly taken into account. Under an imposed shear $\gamma$, every particle tends to reach its affine position $\underline{r}_i^{A}=\underline{\underline{\gamma}} \underline{R}_{j}$ in space, which is entirely specified by the applied strain tensor $\underline{\underline{\gamma}}$ and by the initial position of the particle $\underline{R}_{i}$ in the glass at rest. However, due to the lack of a local center-inversion symmetry in the glass, the forces transmitted to particle $i$ by its nearest-neighbours $j$ cannot just vanish by mutual cancellation with their mirror-images across $i$, as they would do in a lattice with center-inversion symmetry. As a consequence, a net non-zero force $\underline{f}_{i}$ acting on particle $i$ in its affine position $\underline{r}_i^{A}$ pushes particle $i$ towards its final \textit{nonaffine} position, where the particle is, eventually, at mechanical equilibrium. The final equilibrium position can be written as $\underline{r}_i(\gamma_{0})=\underline{r}_i^{A}+\underline{u}_i^{NA}$, where $\underline{u}_i^{NA}$ denotes the nonaffine displacement, or the distance vector between the final equilibrium position of the particle and its affine position. It has been shown that for sufficiently small strains, the nonaffine displacement is also linear in the strain, $\underline{u}_i^{NA}\sim\gamma_{0}$.

Hence, the affine positions $\underline{r}_i^{A}$ of the atoms define a \emph{nonequilibrium} state of the solid. The nonaffine motions represent the trajectory in phase space of the system traveling towards the final equilibrium state. The latter is achieved when all the atoms have reached their final true positions $\underline{r}_i$. The energy associated with the relaxation to equilibrium is given by the entropy change $-T\Delta S$ that corresponds to work dissipated by the system. $\Delta S$ is the entropy difference between the final equilibrium state and the initial nonequilibrium state. From statistical mechanics it is well known that this energy, if the deviation from equilibrium is small (which may be reasonable for $\gamma$ small), coincides with the minimum work $R_{\mathrm{min}}$ that an external agent would do to bring the system from the equilibrium state to the nonequilibrium state that we are considering: $\Delta S=-R_{\mathrm{min}}/T$~\cite{Landau_Stat}. This equality applies to a closed system (the solid) where locally a small body (the particle) moves to reach equilibrium with the system. The total dissipated work can thus be calculated as follows.

We can write the generic force increment $\delta\underline{f}_{i}$ needed to infinitesimally displace the particle from the final equilibrium position. This force, most generally, can be written as a first-order Taylor expansion around the true position in the elastic potential of the nearest-neighbors:
\begin{equation}
\delta\underline{f}_{i}%(\underline{r}_{i}(\underline{\underline{\eta}}))
\,\approx\,-\left(\frac{\partial^2
\mathcal F}{\partial \underline{r}_i\partial
\underline{r}_j}\right)_{\underline{r}_{i}(\gamma)}\,  \cdot\underline{x}_j
=\,-\underline{\underline{H}}_{ij}\, \underline{x}_j \label{eqmod8}
\end{equation}
where $\underline{x}_{j}$ measures the position of each atom $j$ along the nonaffine displacement pathway, and is such that $\underline{x}_{j}=0$ when the atom $j$ is in the true final position $\underline{r}_j$ and $\underline{x}_{j}=\underline{u}_{j}^{NA}$ when the atom is in the affine position $\underline{r}_j^{A}$. According to this definition, the coordinate $\underline{x}_{j}$ measures the distance, along the nonaffine path, that separates the particle $j$ from its true final position $\underline{r}_j$ at equilibrium. $\underline{\underline{H}}_{ij}=\left(\partial^2 E/\partial
\underline{r}_{ij}\partial \underline{r}_{kl}\right)_{\underline{R}_{ij}}$ represents the standard dynamical matrix, or Hessian matrix, of the solid.

Using this expansion we can evaluate $R_{\mathrm{min}}$ defined as the work that would be necessary to bring the particle $i$ from the true nonaffine position (where it is at equilibrium) back to the affine position (where it out of equilibrium):
\begin{equation}
R_{\mathrm{min}}=\int_0^{\underline{u}^{NA} }\,\delta\underline{f}_{i}\cdot \mathrm{d} \underline{x}_i=-\frac{1}{2}\underline{\underline{H}}_{ij} \underline{u}_{i}^{NA} \underline{u}_{j}^{NA}.
\label{eq:w}
\end{equation}
This work is negative because it is the work that an external agent does onto the system, in agreement with the thermodynamic definition of $R_{\mathrm{min}}$.
Using $\Delta S=-R_{\mathrm{min}}/T$, this establishes that the total free energy of the deformed solid, including the contributions from the nonaffine motions, is given by:
\begin{equation}\label{eq:fa}
\mathcal{F}=\mathcal{F}_{A}-T \Delta S=\mathcal{F}_{A}-\frac{1}{2}\underline{\underline{H}}_{ij} \underline{u}_{i}^{NA} \underline{u}_{j}^{NA}
\end{equation}
where $\mathcal{F}_{A}=\frac{1}{8}\left(\partial^2 E/\partial
\underline{r}_{ij}\partial \underline{r}_{kl}\right)
\underline{u}_{ij}^{A} \underline{u}_{kl}^{A}$ denotes the standard Born-Huang free energy of affine deformation for harmonic lattices (the same applies to lattices with inversion symmetry). In earlier independent contributions, it was shown that upon differentiating the total free energy twice with respect to the strain one obtains the following formula for the shear modulus: $G=G^{A}-G_{NA}=G^{A}-\underline{\Xi}_i(\underline{\underline{H}}_{ij})^{-1}\underline{\Xi}_j$, where $\underline{\Xi}_i=\partial \underline{f}_{i}/\partial\gamma_{0}$ denotes the net force per unit strain acting on the particle $i$ resulting from its nearest-neighbour forces. The nonaffine contribution to the shear modulus is thus intimately related to the work $R_{\mathrm{min}}$ and to the entropy associated with degrees of freedom of nonaffine motion, $\Delta S$. It is important to note that for hard-sphere colloids, the usual temperature should be replaced with the the Maxwell relation as $T=(\partial S/\partial E)^{-1}$, where $E$ is the total internal energy of the system.

%Furthermore, we shall follow Prigogine's labelling of the dissipative part of the free energy, by introducing the decomposition of the entropy as $dS=d_{e}S+d_{i}S$, where, in Prigogine's notation, $Td_{e}S$ denotes the reversible part of the entropic contribution to the free energy, while $Td_{i}S$ denotes the irreversible part.

At this point, we can introduce the dissipative function $\Psi_{NA}$ associated with the nonaffine motion. The dissipative function is defined in terms of the entropy production rate $\dot{S}$ by the relation: $2\Psi_{NA}=\dot{S}=(dS/d\underline{x}_{i})\cdot (d\underline{x}_{i}/dt)$. Using $(\partial S/\partial E)^{-1}dS=-dR_{\mathrm{min}}$, and Eq. $dS/d\underline{x}_{i}=\frac{1}{2}(\partial S/\partial E)\underline{\underline{H}}_{ij}\underline{x}_{j}$ from the analysis above, we obtain the following form of the dissipative function,
\begin{equation}
\Psi_{NA}=\frac{1}{2(\partial S/\partial E)^{-1}}\underline{\underline{H}}_{ij}\underline{x}_{j} \dot{\underline{x}}_{i}
\end{equation}
Interestingly, in contrast to the standard viscous dissipation of liquids (and solids), given by $\Psi=\frac{1}{2}\eta\dot{\gamma_{0}}^{2}$ the dissipation function for nonaffine motions is not a quadratic form of the strain rate $\dot{\underline{x}}_{i}$, but just linear.
For sufficiently small strain,
\begin{equation}
\Psi_{NA}\sim \underline{x}_{j} \dot{\underline{x}}_{i} = B \mid\gamma\mid \dot{\gamma_{0}},
\end{equation}
which represents the nonaffine contribution to the total dissipation function in Eq.(2) in the main article.

Also of interest is the fact that the contribution of nonaffine dissipation to the dissipative or viscous stress in the system, is
$\sigma'_{NA}=
\partial\Psi_{NA}/\partial\dot{\underline{x}}_{i}=\frac{1}{2(\partial S/\partial E)^{-1}}\underline{\underline{H}}_{ij}\underline{x}_{j}\propto\gamma_{0}$,
in contrast to the standard viscous stress which is $\sim\dot{\gamma_{0}}$. Importantly, the dissipative contribution of nonaffine motions is independent of the internal friction, unlike the standard viscous dissipation. This outcome is in agreement with recent simulations of shock-wave propagation in frictionless jammed packings where dissipation emerges from the nonaffine motions of the particles in the absence of internal friction.
The stress contribution of nonaffine dissipation to the viscous part of the stress should not be confused with the nonaffine contribution to the elastic stress, which bears a negative sign and decreases the overall elastic stress, as shown by different authors in previous publications.
In general, the total dissipative function can be written as the sum of the standard viscous dissipative function and nonaffine one, as
$\Psi_{tot}=\Psi+\Psi_{NA}$. In the solid-like linear regime of deformation $\gamma_{0}<\gamma_{0}^{*}$, $\Psi_{tot}=\Psi+\Psi_{NA}$ with both contributions, while in the liquid-like regime post-yielding $\gamma_{0}>\gamma_{0}^{*}$, we have $\Psi_{tot}=\Psi$, because $\Psi_{NA}$ is defined only for nonaffine deformations of a solid.
It is difficult to quantitatively evaluate the nonaffine contribution $\Psi_{NA}$, in comparison with the standard one. In highly viscous systems like colloidal suspensions, it may be the case that $\Psi_{NA}\ll\Psi$ is a possible outcome. However, we can at least extract scaling laws with the present theory, and analyse different limits.

For example, the maximum value of the dissipation function is achieved in the affine positions: $\Psi_{\mathrm{max}}=(\partial S/\partial E)(\underline{\Xi}_{i}\cdot\underline{\dot{x}}_{i}|_{\underline{r}^{A}_{i}})\gamma_{0}$, where $\underline{\dot{x}}_{i}|_{\underline{r}^{A}_{i}}$ denotes the initial velocity of particle $i$ when it is in its affine position $\underline{r}^{A}_{i}$ and begins its motion towards the final position. In a crystal lattice, where $\underline{\Xi}_{i}=0$ for all atoms $i$ under quasistatic deformation, this implies that $\Psi=0$, at any point in the deformation. Hence, nonaffine dissipation is zero only in the quasistatic deformation of ordered lattices.

We can also analyse the behaviour of $\Psi_{NA}$ as a function of the frequency, or equivalently, rate of deformation.
At very high rates or frequencies, the particles have no time to relax from the affine positions. As is well known, high-frequency or high rates experiments probe the affine deformation and one can assume $\underline{x}_{i}=0$ for all particles, in this limit. Hence, $\Psi_{NA}\rightarrow0$, in the limit $\omega\rightarrow\infty$ and $\dot{\gamma_{0}}\rightarrow\infty$.
In the opposite limit of quasistatic deformation at zero-frequency and zero-strain rate, we have that $\underline{x}_{i}>0$, because particles can relax and undergo nonaffine displacements on the time scale of deformation, but now $\dot{\underline{x}}_{i}=0$.
Hence, $\Psi_{NA}\rightarrow0$ also in the limit $\omega\rightarrow0$ and $\dot{\gamma_{0}}\rightarrow0$.
We conclude that $\Psi_{NA}\neq0$ only at intermediate rates and frequencies.


\begin{thebibliography}{99}
\bibitem{ESRF} Additional experiments have been carried out at the DUBBLE beamline of the ESRF synchrotron facility. The experimental conditions have been similar to the ones performed at DESY, and the rheometer used identical to the one at DESY.
\end{thebibliography}
\end{document}